\documentclass[9pt,twocolumn,twoside]{opticajnl}
\journal{opticajournal} 

\setboolean{shortarticle}{true}

\usepackage{lineno}

\usepackage{dsfont}
\usepackage{mathtext}

\usepackage{mathtools}
\usepackage{amsmath}
\usepackage{amssymb,mathrsfs}
\usepackage{graphicx}
\usepackage{epstopdf}
\usepackage{yfonts}
\usepackage{bm}

\usepackage{color}

\newcommand{\br}{{\bm r}}
\newcommand{\bk}{{\bm k}}

\title{Vortex solitons in moir\'{e} optical lattices}

\author[1,*]{Sergey K. Ivanov}
\author[2]{Vladimir V. Konotop}
\author[3]{Yaroslav V. Kartashov}
\author[1,4]{Llu\'{i}s Torner}

\affil[1]{ICFO-Institut de Ciencies Fotoniques, The Barcelona Institute of Science and Technology, 08860 Castelldefels (Barcelona), Spain}
\affil[2]{Departamento de F\'isica and Centro de F\'isica Te\'orica e Computacional, Faculdade de Ci\^encias, Universidade de Lisboa, Campo Grande, Ed. C8, Lisboa 1749-016, Portugal}
\affil[3]{Institute of Spectroscopy, Russian Academy of Sciences, Fizicheskaya Str., 5, Troitsk, Moscow, 108840, Russia}
\affil[4]{Universitat Politecnica de Catalunya 08034 Barcelona, Spain}

\affil[*]{Corresponding author: sergey.ivanov@icfo.eu}

\begin{abstract}
We show that optical moir\'e lattices enable the existence of vortex solitons of different types in self-focusing Kerr media. We address the properties of such states both in lattices having commensurate and incommensurate geometries (i.e., constructed with Pythagorean and non-Pythagorean twist angles, respectively), in the different regimes that occur below and above the localization-delocalization transition. We find that the threshold power required for the formation of vortex solitons strongly depends on the twist angle and, also, that the families of solitons exhibit intervals where their power is a nearly linear function of the propagation constant and they exhibit a strong stability. Also, in the incommensurate phase above the localization-delocalization transition, we found stable embedded vortex solitons whose propagation constants belong to the linear spectral domain of the system.
\end{abstract}

\begin{document}

\maketitle

Optical moir\'{e} lattices (MLs) have been shown to be a versatile tool for controlling and manipulating light propagation. They enable light localization~\cite{Huang-16,Wang-20},  specific reflectivity by metasurfaces~\cite{Hu-20,Hu-21}, magic-angle lasers~\cite{Mao-21}, and can be used to create flat-bands~\cite{Huang-16,Wang-20} and topologically nontrivial structures~\cite{Lou-21,Dong-21,Yi-21}. Both mono-layered and bi-layered MLs, are formed by the overlap of two identical sublattices rotated with respect to each other. Depending on the twist angle, an ML is either a commensurate (i.e., periodic) or an incommensurate (i.e., aperiodic) structure. Importantly, in both cases, MLs inherit the rotational symmetry of the sublattices. The spectral changes inherent to the transition between commensurate and incommensurate configurations are responsible for rich physical phenomena, including the occurrence of a localization-delocalization transition (LDT) of light beams~\cite{Huang-16,Wang-20}. In nonlinear media, MLs significantly affect the properties of existing soliton families. In particular, optical moir\'{e} lattices can support thresholdless two-dimensional (2D) Kerr solitons~\cite{Fu-20}, while multifrequency solitons in quadratic media acquire properties that are unusual for translationally periodic lattices~\cite{Kartashov-21}. In this context, the impact of incommensurability on the potential existence and properties of vortex solitons (VSs) remains unexplored to date. 

In this Letter, we predict that mono-layered MLs may sustain stable VSs and that the localization properties of the linear spatial eigenmodes of MLs have a strong impact on the power threshold necessary for the formation of the solitons. In the incommensurate phase, MLs enable the excitation of thresholdless VSs that remain spatially localized even in the low-amplitude limit. Furthermore, we find that stability domains for VSs occur above the LDT threshold.

\begin{figure*}[t!]
\centering
\includegraphics[width=\linewidth]{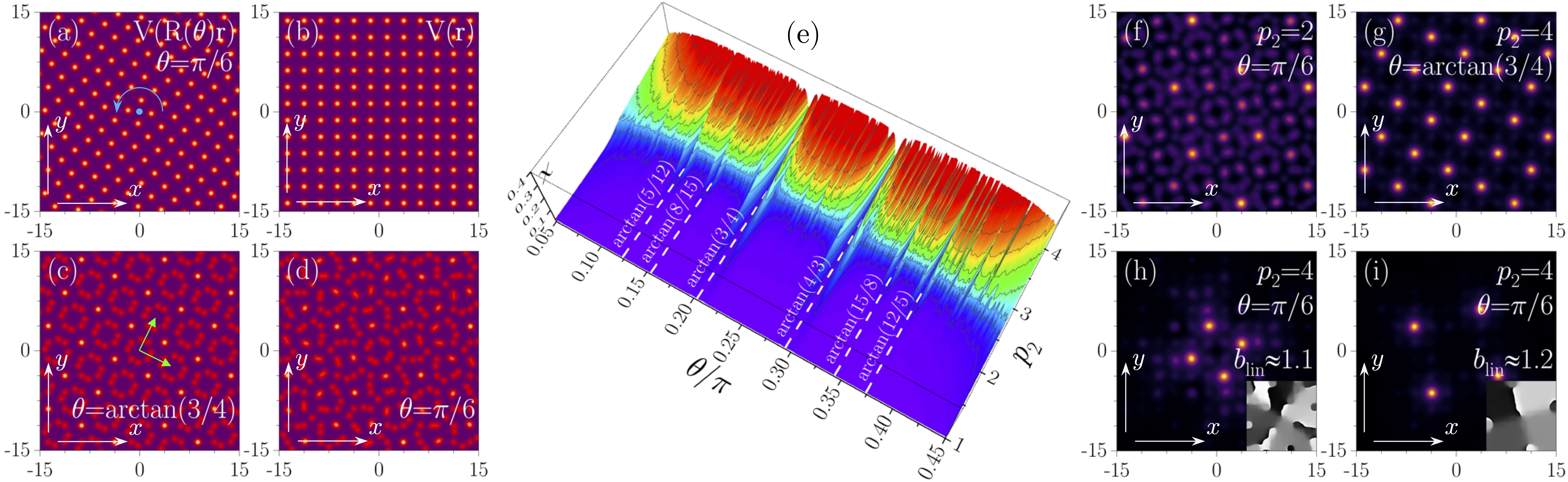}
\caption{
Profiles of the sublattices $V[R(\theta)r]$ with $\theta=\pi/6$ (a) and $V(r)$ (b). In panel (a) the blue dot and arrow indicate the axis and the direction of rotation. Examples of commensurate (c) and incommensurate (d) moir\'{e} lattice $\mathcal{P}(\br)$ produced by the square sublattices with $p_2=2$ rotated by angles $\theta=\pi/6$ (c) and $\theta=\arctan{(3/4)}$ (d). The green arrows in (c) indicate the primitive lattice vectors. (e) Form-factor $\chi$ of the eigenmode with the largest propagation constant versus $\theta$ and $p_2$. White dashed lines indicate some of the Pythagorean angles. Linear modes supported by the moir\'{e} lattices  below (f) and above (g)--(i) LDT threshold. Panel (g) corresponds to a commensurate lattice, while (f), (h), and (i) correspond to an incommensurate one. Linear vortex modes $|\psi|$, and $\arg(\psi)$ shown in the insets, have $b_{\mathrm{lin}}=1.061$ (h) and $b_{\mathrm{lin}}=1.207$ (i) with $M=1$. Here and below $p_1=2$, $a=2.5$ and $\sigma=0.5$.
}
\label{fig1}
\end{figure*}

We consider the propagation of paraxial light beams in a nonlinear cubic medium with a transverse shallow modulation of the refractive index that has the form of a Pythagorean ML, which is described by the nonlinear Schr\"{o}dinger equation for the dimensionless light field amplitude $\psi$
\begin{equation} 
\label{NLS}
    i \frac{\partial \psi}{\partial z} =  H_0\psi - |\psi|^2\psi, \quad H_0=-\frac{1}{2} \nabla^2  - \mathcal{P}(\br).
\end{equation}
Here $\br=(x,y)$ and $\nabla=(\partial_x,\partial_y)$. The transverse coordinates ${\br}$ and the propagation distance $z$ are normalized to the characteristic transverse scale $w$ and the diffraction length $\kappa w^2$, respectively, where $\kappa$ is the wavenumber. The function $\mathcal{P}(\br)=p_1V(R(\theta)\br) + p_2V(\br)$ describes MLs created by the superposition of two square sublattices $V(\br) = \sum_{m,n}Q\left(x-na+a/2,y-ma+a/2\right)$ with $Q(x,y)=\exp\left[-\left(x^2+y^2\right)/\sigma^2\right]$ and different waveguide depths $p_1$ and $p_2$ corresponding to the refractive index contrast $\delta n$, $p=\kappa^2 w^2 \delta n/n_0$, where $n_0$ is the refractive index, $R(\theta)$ is the operator of 2D rotation by the angle $\theta$. In the numerical calculations reported below, each sublattice is set to have the period $a=2.5$ and to be composed of identical waveguides of width $\sigma=0.5$ [Figs.~\ref{fig1}(a) and (b)]. To each Pythagorean angle, defined by $\theta=\arctan[(m^2-n^2)/2mn]$, with  $m>n>0$, and  $m$, $n \in \mathbb{N}$, one can associate the Pythagorean triple $(m^2-n^2,2mn,m^2+n^2)$. When the twist angle is a Pythagorean one, the ML is periodic (commensurate) [Fig.~\ref{fig1}(c)]. Otherwise, the array is aperiodic (incommensurate) [Fig.~\ref{fig1}(d)]. Such MLs can be induced in suitable photosensitive materials~\cite{Wang-20,Fu-20} or inscribed with fs laser pulses in transparent dielectrics~\cite{Arkhipova-23}.

The properties of the linear modes supported by the MLs may change qualitatively not only due to the change of the twist angle but also when the depths of the sublattices vary~\cite{Huang-16,Wang-20}. Therefore, we first calculate the eigenmodes of the linear Hamiltonian $H_0$. Such  linear modes can be written in the form $\psi(\br,z)=u(\br) e^{ib_{\mathrm{lin}}z}$, where $b_{\mathrm{lin}}$ is the eigenvalue, i.e., the propagation constant of the mode and the function $u(\br)$ describes the transverse mode profile. {{To calculate the spectrum and modes of the system we used a standard finite difference method (see Supplement).} The width of the modes is characterized by the inverse form-factor $\chi^{-1}$, where $\chi=U^{-1}(\int|\psi|^4d\br)^{1/2}$ and $U=\int|\psi|^2d\br$ is the power. High $\chi$ stands for strong localization. The form-factor of the linear mode with the largest propagation constant is plotted in Fig.~\ref{fig1}(e) as a function of $\theta$ and $p_2$ at a fixed $p_1=2$. For Pythagorean angles - some of which correspond to the dashed lines - all modes are delocalized, as they  are Bloch modes. For non-Pythagorean angles, LDT is encountered when $p_2$ reaches a certain threshold value $p_2^{cr}\approx 3$. Illustrative examples of linear modes for various parameters are shown in Figs.~\ref{fig1}(f)-(i).

The point-symmetry group of the Pythagorean ML considered here is the dihedral D$_4$ group for all twist angles, except for $\theta=\pi/4$ and $p_1=p_2$, when it acquires the higher symmetry D$_8$. Thus, for $\theta\neq\pi/4$  it can support only states with vorticity $M=0,\pm 1$, while, in agreement with fundamental selection rules, for $\theta=\pi/4$ and $p_1=p_2$ vortices with $M=\pm2,\,\pm 3$ are allowed too~\cite{Ferrando-05(1)}. The reflection symmetries allow restricting the consideration to angles $\theta\in[0,\pi/4]$. They also explain the symmetry of the form-factor in Fig.~\ref{fig1} (e) with respect to $\theta=\pi/4$, as well as that the properties of states with equal absolute values, but different signs of topological charges, are identical. Examples of the modulus and phase distributions of localized linear vortex modes for an incommensurate ML above the LDT are shown in Fig.~\ref{fig1}(h) and~(i). Such states are constructed using superposition of two conventional linear modes of ML at $\theta=\pi/6$ and $p_2=4$. The more compact the vortex the smaller propagation constant [c.f.\ panel~(h) with panel~(i) in Fig.~\ref{fig1}].

\begin{figure*}[t!]
\centering
\includegraphics[width=\linewidth]{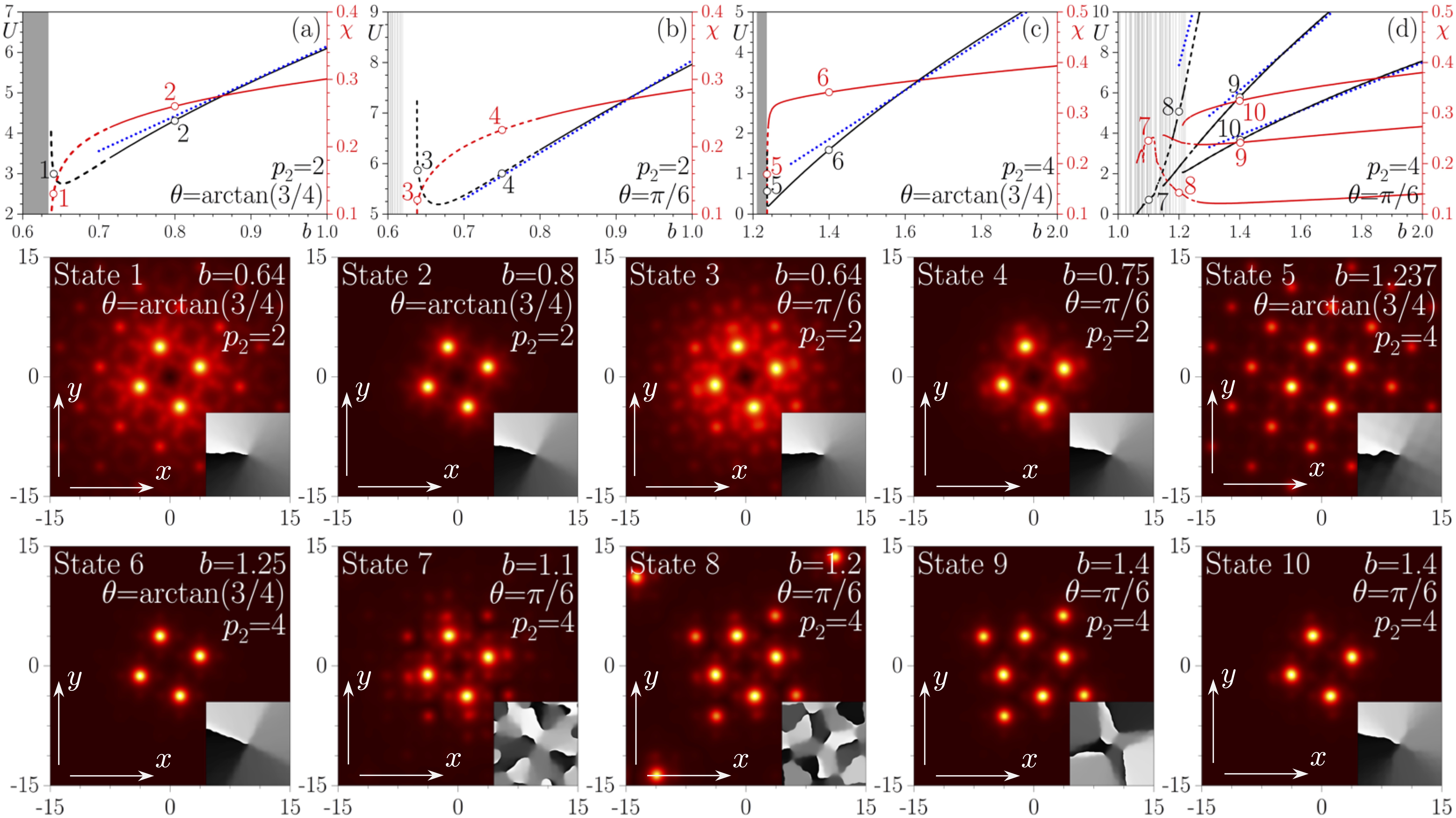}
\caption{
Power $U(b)$ (black curves) and  form-factor $\chi$ (red curves) of vortex solitons versus propagation constant in moir\'{e} lattices with $p_2=2$, $\theta=\arctan{(3/4)}$ (a) and $\theta=\pi/6$ (b) and $p_2=4$, $\theta=\arctan{(3/4)}$ (c) and $\theta=\pi/6$ (d). Solid and dashed lines correspond to stable and unstable families. Shaded regions show allowed bands for commensurate lattices, while vertical gray lines indicate propagation constants of the linear states of incommensurate lattices. {{The blue dotted lines depict the approximation of $U(b)$ by the linear dependence $U\propto \left(b-b^{\rm co}\right)/\chi^2(b)$} (see the text)}. Open circles with numbers correspond to vortex solitons, whose profiles ($|\psi|$ distributions and $\arg(\psi)$ distributions in the insets) are  depicted in the middle and bottom rows.
}
\label{fig2}
\end{figure*}

We now address VSs solutions in Eq.~(\ref{NLS}) with included cubic nonlinearity. It is known that, in general, vortex states can be stabilized by lattices that suppress azimuthal modulation instabilities (see reviews~\cite{Desyatnikov-05,Pryamikov-21}, theoretical predictions~\cite{Malomed-01,Yang-03,Alexander-04,Oster-06}, and experimental observations~\cite{Neshev-04,Fleischer-04,Terhalle-08}). {{The stability of VSs of the gap type, supported by a quasiperiodic lattice, was examined in~\cite{Sakaguchi-06}.}} Structures with discrete angular rotation symmetry (without transverse periodicity) also support vortex solitons~\cite{Kartashov-04(1),Wang-06,Fischer-06,Ferrando-05(2)}. However, to date, the existence and stability of VSs has not been explored in MLs. Here we found that the stability of VSs and their properties are connected with the characteristics of the linear eigenmodes of MLs analyzed above that undergo LDT. 

To this end, we searched for soliton solutions in the form $\psi(\br,z)=u(\br) e^{ibz}$, where $b$ is a propagation constant, and $u(\br)$ is a complex function defining the transverse profile. The properties of such solutions are summarized in Fig.~\ref{fig2}. The top row displays the dependencies of the power $U$ and form-factor $\chi$ on $b$ for both Pythagorean $\theta=\arctan{(3/4)}$ and non-Pythagorean $\theta=\pi/6$ angles. The gray areas correspond to the allowed bands for the Pythagorean angle and the gray vertical lines correspond to the propagation constants $b_{\mathrm{lin}}$ of the linear modes obtained by numerical calculation, where we considered a large, but finite, window, for the non-Pythagorean angle. In commensurate lattices, VSs exist above the cutoff propagation constant $b^{\rm co}$ such that $b_{\rm lin}<b^{\rm co}$. We found the existence of the threshold power $U_{\rm th}$ below which VSs cannot exist, a results that manifest itself by the non-monotonous dependence $U(b)\geq U_{\rm th}$ [black lines in Figs.~\ref{fig2} (a) and (c)]. A similar phenomenon occurs also in  incommensurate lattices below the LDT [Fig.~\ref{fig2}(b)]. In all these cases, the form-factor monotonically decreases as the propagation constant approaches the cutoff, consistent with the delocalization of the VSs near it. Examples of the respective VSs are shown in the middle and bottom rows of Fig.~\ref{fig2} (States 1 to 6). 

However, like in the case of fundamental solitons~\cite{Fu-20}, a very different behavior is encountered in the incommensurate case generated by non-Pythagorean angles, above the LDT [Fig.~\ref{fig2}(d)]. Now, the families of VSs bifurcate from linear vortex modes that may lie deep in the linear spectrum, thus they become thresholdless. As a consequence, the respective nonlinear solutions can be viewed as {\em embedded VSs}, by analogy with the similar concept known for other types of  solitons~\cite{Buryak1995,Yang1999}). For example, the family shown in Fig.~\ref{fig2}(d) bifurcates from the mode shown in Fig.~\ref{fig1}(h) with $b_{\mathrm{lin}}=1.061$.
An example of a nonlinear vortex state near the bifurcation point is shown in the bottom row of Fig.~\ref{fig2} (State~7). Unlike the nonlinear states that exist near the cutoff in commensurate MLs, this state is spatially localized. When $b$ increases we observe two phenomena. First, the width of the VS belonging to the same family as State~7 can be bigger for larger propagation constants (cf. States 7 and 8). Second, new families of solutions arise (States 9 and 10). They emerge around propagation constant values that coincide with those of the linear states having spatial distributions close to the distribution of the nonlinear modes. For instance, State~9 arises around the linear vortex with $b_{\mathrm{lin}}=1.207$ shown in Fig.~\ref{fig1}(i). This leads to the rather complex bifurcation structure depicted in Fig.~\ref{fig2}(d), where only the main VS families are shown. For all such families the power $U$ monotonically increases with $b$.
{{As mentioned above, under specific parameter configurations, the moir\'{e} array has the capability to sustain states with $M=\pm 2,\,\pm 3$. However, we found that such solutions are usually unstable over a wide range of propagation constants (see Supplement).}}
 
A noticeable property is found to occur for propagation constants sufficiently detuned from the cutoff value: The $U(b)$ dependence in this regime is well approximated by the linear dependence $U\propto \left(b-b^{\rm co}\right)/\chi^2(b)$, shown by the blue dotted lines in the top row of Fig.~\ref{fig2}, where the mean value of the curve slope $\chi(b)$ was obtained from direct numerical calculations. This phenomenon can be understood as a result of the flat bands exhibited by MLs. Namely, on the one hand, even in the incommensurate case, for a localized solution, an ML can be approximated by a Pythagorean periodic lattice~\cite{Wang-20}. A feature of such approximation is that it has an extremely large primitive cell - and the larger the cell the better the approximation. The relative flatness of the bands occurs also for commensurate lattices. On the other hand, states with relatively small amplitude can be searched in the form of the multiscale expansion $\psi\approx A(\br,t)e^{ib^{\rm co}z}$ where $A(\br,t)$ is the amplitude slowly varying in space and in time. In general, $A(\br,t)$ solves the 2D nonlinear Schr\"odinger equation~\cite{Baizakov2002} with the effective dispersion coefficients $D_{x,y}\propto \partial^2b_\alpha(\bk)/\partial k_{x,y}^2$ ($\alpha$ is the number of a linear band above which the soliton family is located) and the effective nonlinear coefficient $\chi_{\rm eff}=U^{-1}(\int |\psi_\alpha^{\rm co}|^4d\br)^{1/2}$. In the case of nearly flat bands, when the dispersion is suppressed, i.e., when $D_x=D_y=0$ in the leading order, the shape of the beam becomes weakly dependent on the amplitude allowing one to approximate $\chi_{\rm eff}\approx \chi$ (where $\chi$ is the form-factor defined above). Then the equation describing the amplitude is reduced to $i\partial_t A=\chi |A|^2 A$. Averaging this equation over the transverse direction and using that $U\approx \int |A|^2 d\br$, yields the estimate $b=b^{\rm co}+\chi^2 U$, from which we obtained the above mentioned linear dependence.

We use the standard linear stability analysis to elucidate the stability of the VSs. Thus, perturbed solutions are substituted into Eq.~(\ref{NLS}), a  linearization around the stationary state is performed, and the solution of the resulting linear eigenvalue problem yields the perturbation growth $\lambda$ (see Supplement for the details and $\lambda(b)$ dependencies). Solitons are stable when Re$\lambda=0$. The results of the analysis for the explored families of VSs are shown by solid (stable) and dashed (unstable) lines in the upper row of Fig.~\ref{fig2}. Below the LDT, VSs are unstable in the vicinity of the cut-off propagation constant $b^{\rm co}$ [panels (a) and (b)], while relatively large stability intervals appear at larger propagation constants, where the families $U(b)$ are well approximated by the linear law. Such stability can be understood as a consequence of the flat-bands exhibited by the MLs, which weaken diffraction. This understanding corroborates with enhanced stability above LDT,  both in a commensurate phase [nearly in the entire interval of $b>b^{\rm co}$ in Fig.~\ref{fig2} (c)] and in an incommensurate phase for the low-energy modes [Fig.~\ref{fig2} (d)], where stable families are found even with propagation constants embedded in the linear spectrum (see States 7 and 8). {Notice the enlarged stability regions of vortex states at large $b$ values due to the increased localization of the spots forming such states,leading to weaker interactions between them.}

\begin{figure}[t]
\centering\includegraphics[width=\linewidth]{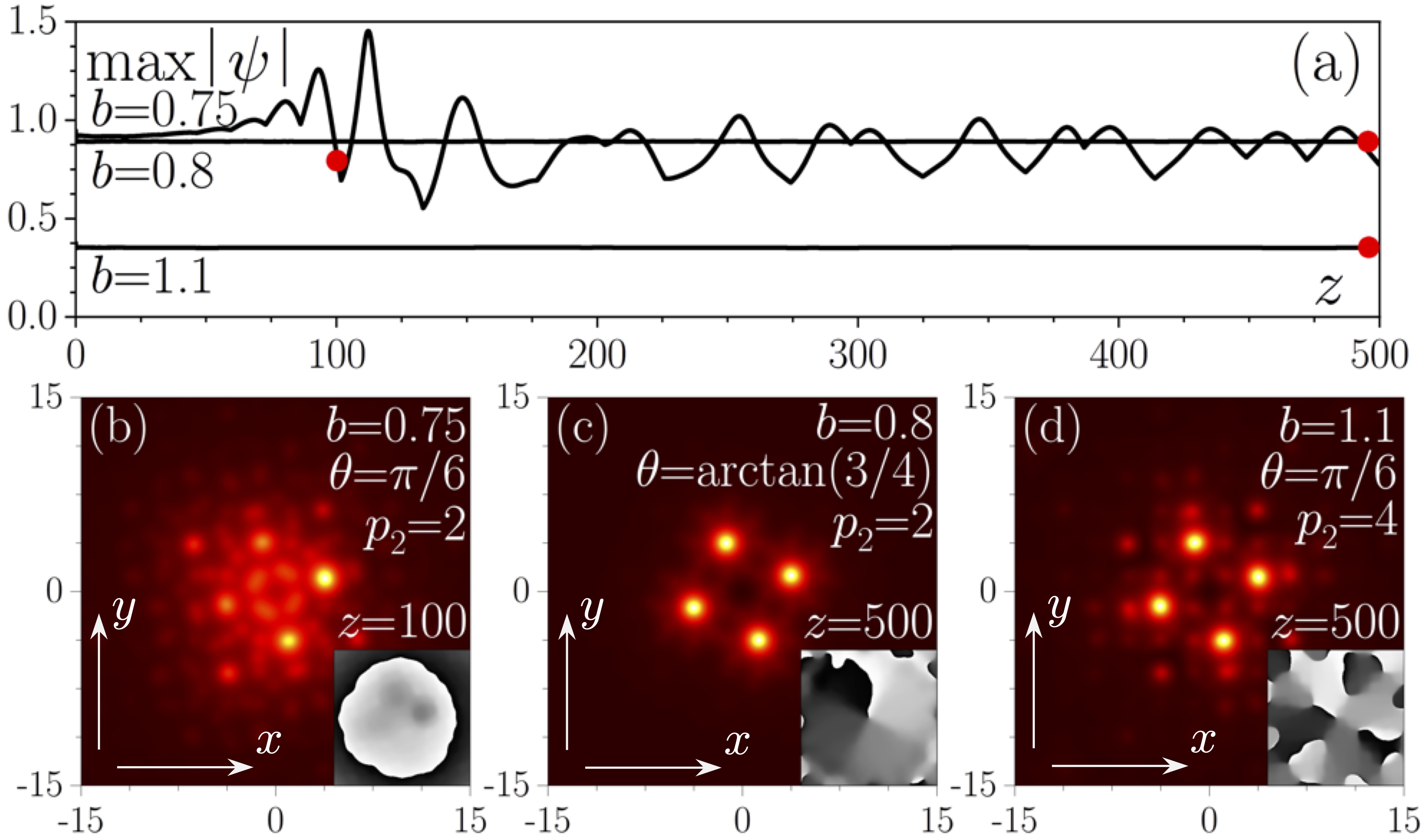}
\caption{
Propagation of VSs. The top row shows the dependence of the peak amplitudes on distance, while the bottom row shows profiles of unstable states with $b=0.75$, $\theta=\pi/6$, $p_2=2$ (b) at $z=100$ and profiles of stable vortex solitons with $b=0.8$, $\theta=\arctan{(3/4)}$, $p_2=2$ (c) and $b=1.1$, $\theta=\pi/6$, $p_2=4$ (d) at $z=500$. The red dots on panel (a) correspond to the instants for which distributions in (b)-(d) are shown.
}
\label{fig3}
\end{figure}

Finally, in Fig.~\ref{fig3} we show examples of the dynamical evolution of VSs. The upper panel (a) illustrates the evolution of the amplitudes of two stable VSs in commensurate ($b=0.8$) and in incommensurate ($b=1.1$) MLs, {{ as well as one unstable and rapidly decaying VS}} ($b=0.75$) in an incommensurate ML. The respective input ($z=0$) field distributions are shown in Fig.~\ref{fig2} (see States~2, 4, and 7). Panel (b) in Fig.~\ref{fig3} shows a snapshot of the unstable soliton at a distance $z=100$ (see also Supplement), while panels~(c) and~(d) show the snapshots of stable VSs at $z=500$. The distances at which $|\psi|$ distributions are shown are indicated by the red dots in Fig.~\ref{fig3}(a).


In summary, we have shown that the existence, stationary properties, and dynamical stability of vortex solitons in moir\'e optical lattices strongly depend on whether the lattice is commensurate or incommensurate and, also, on whether the lattice parameters are below or above the LDT. The families of nonlinear modes feature a linear dependence of the energy versus the propagation constant detuning for a wide range of propagation constants. The analyzed vortex solitons exhibit strong stability. Also, when the twist angle is non-Pythagorean, the moir\'{e} lattices sustain stable embedded vortex solitons.

\medskip

\begin{backmatter}
\bmsection{Funding} 

ICFO was supported by Agencia Estatal de Investigación (CEX2019-000910-S, PGC2018-097035-B-I00), Generalitat de Catalunya (47Y48R6YK), CERCA, Fundació Cellex, and Fundació Mir-Puig. V.V.K. was supported by the Portuguese Foundation for Science and Technology (PTDC/FIS-OUT/3882/2020 and UIDB/00618/2020). Y.V.K.’s academic research was supported by Russian Science Foundation (grant 21-12-00096) and research project FFUU-2021-0003 of the Institute of Spectroscopy of the Russian Academy of Sciences.

\bmsection{Disclosures}
The authors declare no conflicts of interest.

\bmsection{Data availability}
All data underlying the results presented in this paper may be obtained from the authors upon reasonable request.

\bmsection{Supplemental document}
See Supplement for supporting content. 

\end{backmatter}



\begin{thebibliography}{99}



\bibitem{Huang-16}
C. Huang, F. Ye, X. Chen, Y. V. Kartashov, V. V. Konotop, and L. Torner,
``Localization-delocalization wavepacket transition in Pythagorean aperiodic potentials,''
Sci. Rep. \textbf{6}, 32546 (2016).

\bibitem{Wang-20}
P. Wang, Y. Zheng, X. Chen, C. Huang, Y. V. Kartashov, L. Torner, V. V. Konotop, and F. Ye,
``Localization and delocalization of light in photonic moir\'{e} lattices,''
Nature \textbf{577}, 42 (2020).

\bibitem{Hu-20}
 G. W. Hu, A. Krasnok, Y. Mazor, C. W. Qu, and A. Al\`{u},
 ``Moir\'{e} hyperbolic metasurfaces,''
 Nano Lett. \textbf{20}, 3217 (2020).

\bibitem{Hu-21}
G. Hu, M. Wang, Y. Mazor, C.-W. Qiu, and A. Al\`{u},
``Tailoring light with layered and moir\'{e} metasurfaces,''
Trends Chem. \textbf{3}, 342 (2021).

\bibitem{Mao-21}
X.-R. Mao, Z.-K. Shao, H.-Y. Luan, S.-L. Wang, and R.-M. Ma,
``Magic-angle lasers in nanostructured moir\'{e} superlattice,''
Nat. Nanotechnol. \textbf{16}, 1099 (2021).

\bibitem{Lou-21}
B. Lou, N. Zhao, M. Minkov, C. Guo, M. Orenstein, and S. Fan,
``Theory for twisted bilayer photonic crystal slabs,''
Phys. Rev. Lett. \textbf{126}, 136101 (2021).

\bibitem{Dong-21}
K. Dong, T. Zhang, J. Li, Q. Wang, F. Yang, Y. Rho, D. Wang, C. P. Grigoropoulos, J. Wu, and J. Yao,
``Flat bands in magic-angle bilayer photonic crystals at small twists,''
Phys. Rev. Lett. \textbf{126}, 223601 (2021).

\bibitem{Yi-21}
C.-H. Yi, H. C. Park, and M. J. Park,
``Strong interlayer coupling and stable topological flat bands in twisted bilayer photonic Moiré superlattices,''
Light: Sci.\& App.  {\bf 11}, 289 (2022) 


\bibitem{Fu-20}
Q. Fu, P. Wang, C. Huang, Y. V. Kartashov, L. Torner, V. V. Konotop, and F. Ye,
``Optical soliton formation controlled by angle twisting in photonic moir\'{e} lattices,''
Nat. Photonics \textbf{14}, 663 (2020).

\bibitem{Kartashov-21}
Y. V. Kartashov, F. Ye, V. V. Konotop, and L. Torner,
``Multifrequency solitons in commensurate-incommensurate photonic Moir\'{e} lattices,''
Phys. Rev. Lett. \textbf{127}, 163902 (2021).


\bibitem{Arkhipova-23}
A. A. Arkhipova, Y. V. Kartashov, S. K. Ivanov, S. A. Zhuravitskii, N. N. Skryabin, I. V. Dyakonov, A. A. Kalinkin, S. P. Kulik, V. O. Kompanets, S. V. Chekalin, F. Ye, V. V. Konotop, L. Torner, and V. N. Zadkov,
``Observation of linear and nonlinear light localization at the edges of Moir\'{e} arrays,''
Phys. Rev. Lett. \textbf{130}, 083801 (2023).



\bibitem{Ferrando-05(1)}
A. Ferrando, M. Zacares, and M. A. Garcia-March,
``Vorticity cutoff in nonlinear photonic crystals,''
Phys. Rev. Lett. \textbf{95}, 043901 (2005).


\bibitem{Desyatnikov-05}
A. S. Desyatnikov, Y. S. Kivshar, and L. Torner,
``Optical vortices and vortex solitons,''
Prog. Opt. \textbf{47}, 291 (2005).

\bibitem{Pryamikov-21}
A. Pryamikov, L. Hadzievski, M. Fedoruk, S. Turitsyn, and A. Aceves,
``Optical vortices in waveguides with discrete and continuous rotational symmetry,''
J. Eur. Opt. Soc.-Rapid Publ. \textbf{17}, 23 (2021).


\bibitem{Malomed-01}
B. A. Malomed and P. G. Kevrekidis,
``Discrete vortex solitons,''
Phys. Rev. E \textbf{64}, 026601 (2001).

\bibitem{Yang-03}
J. Yang and Z. H. Musslimani,
``Fundamental and vortex solitons in a two-dimensional optical lattice,''
Opt. Lett. \textbf{28}, 2094 (2003).

\bibitem{Alexander-04}
T. J. Alexander, A. A. Sukhorukov, and Y. S. Kivshar,
``Asymmetric vortex solitons in nonlinear periodic lattices,''
Phys. Rev. Lett. \textbf{93}, 063901 (2004).

\bibitem{Oster-06}
M. \"{O}ster and M. Johansson,
``Stable stationary and quasiperiodic discrete vortex breathers with topological charge $S=2$,''
Phys. Rev. E \textbf{73}, 066608 (2006).


\bibitem{Neshev-04}
D. N. Neshev, T. J. Alexander, E. A. Ostrovskaya, Y. S. Kivshar, H. Martin, I. Makasyuk, and Z. Chen,
``Observation of discrete vortex solitons in optically induced photonic lattices,''
Phys. Rev. Lett. \textbf{92}, 123903 (2004).

\bibitem{Fleischer-04}
J. W. Fleischer, G. Bartal, O. Cohen, O. Manela, M. Segev, J. Hudock, and D. N. Christodoulides, 
``Observation of vortex-ring discrete solitons in 2D photonic lattices,''
Phys. Rev. Lett. \textbf{92}, 123904 (2004).

\bibitem{Terhalle-08}
B. Terhalle, T. Richter, A. S. Desyatnikov, D. N. Neshev, W. Krolikowski, F. Kaiser, C. Denz, and Y. S. Kivshar,
``Observation of multivortex solitons in photonic lattices,''
Phys. Rev. Lett. \textbf{101}, 013903 (2008).


\bibitem{Sakaguchi-06}
H. Sakaguchi and B. A. Malomed,
``Gap solitons in quasiperiodic optical lattices,''
Phys. Rev. E \textbf{74}, 026601 (2006).


\bibitem{Kartashov-04(1)}
Y. V. Kartashov, V. A. Vysloukh, and L. Torner,
``Rotary solitons in Bessel optical lattices,''
Phys. Rev. Lett. \textbf{93}, 093904 (2004).

\bibitem{Wang-06}
X. S. Wang, Z. G. Chen, and P. G. Kevrekidis,
``Observation of discrete solitons and soliton rotation in optically induced periodic ring lattices,''
Phys. Rev. Lett. \textbf{96}, 083904 (2006).

\bibitem{Fischer-06}
R. Fischer, D. N. Neshev, S. Lopez-Aguayo, A. S. Desyatnikov, A. A. Sukhorukov, W. Krolikowski, and Yu. S. Kivshar,
``Observation of light localization in modulated Bessel optical lattices,''
Opt. Express \textbf{14}, 2825 (2006).


\bibitem{Ferrando-05(2)}
A. Ferrando, M. Zacares, M. A. Garcia-March, J. A. Monsoriu, and P. F. de Cordoba,
``Vortex transmutation,''
Phys. Rev. Lett. \textbf{95}, 123901 (2005).


\bibitem{Buryak1995} A. V. Buryak, 
``Stationary soliton bound states existing in resonance with linear waves,''
Phys. Rev. E {\bf 52}, 1156 (1995).

\bibitem{Yang1999} J. Yang, B. A. Malomed, and D. J. Kaup, 
``Embedded Solitons in Second-Harmonic-Generating Systems,''
Phys. Rev. Lett. {\bf 83}, 1958 (1999).

\bibitem{Baizakov2002}
B. B. Baizakov, V. V. Konotop, and M. Salerno, 
``Regular spatial structures in arrays of Bose-Einstein condensates induced by modulational instability,''
J. Phys. B {\bf 35}, 5105 (2002)


\end{thebibliography}

\end{document}